%% file: main.tex
\documentclass[preprint,review,12pt]{elsarticle}

\input{mypackages}
\input{mycommands}

\newcommand\copyrighttext{%
    \footnotesize This work has been submitted to the Elsevier (Pervasive and Mobile Computing Journal) for possible publication. Copyright may be transferred without notice, after which this version may no longer be accessible.}
    \newcommand\copyrightnotice{%
	\begin{tikzpicture}[remember picture,overlay]
	     \node[anchor=south,yshift=10pt] at (current page.south) {\fbox{\parbox{\dimexpr\textwidth-\fboxsep-\fboxrule\relax}{\copyrighttext}}};
	\end{tikzpicture}%
}

\journal{Pervasive and Mobile Computing}

\begin{document}

\begin{frontmatter}

\title{Resilient UAVs Location Sharing Service Based on Information Freshness and Opportunistic Deliveries}

\author[label1]{Agnaldo Batista}
\author[label1,label2]{Aldri L. dos Santos}

\address[label1]{Department of Informatics, Federal University of Paraná, 
Brazil}
\address[label2]{Department of Computer Science, Federal University of Minas Gerais, 
Brazil}

\begin{abstract}
Unmanned aerial vehicles (UAV) have been recognized as a versatile platform for a wide range of services. During the flight, these vehicles must avoid collisions to operate safely. In this way, they demand to keep spatial awareness, i.e., to know others in their coverage area. However, mobility and positioning aspects hamper building UAV network infrastructure to support reliable basic services. Thus, such vehicles call for a location service with up-to-date information resilient to false location injection threats. This work proposes FlySafe, a resilient UAVs location sharing service that employs opportunistic approaches to deliver UAVs' location. FlySafe takes into account the freshness of UAVs' location to maintain their spatial awareness. Further, it counts on the age of the UAV's location information to trigger device discovery. Simulation results showed that FlySafe achieved spatial awareness up to 94.15\% of UAV operations, being resilient to~false locations injected in the network. Moreover, the accuracy in device discovery achieved 94.53\% with a location error of less than 2~m.
\end{abstract}




\end{frontmatter}
\copyrightnotice

\section{Introduction}

Unmanned aerial vehicles (UAV) or drones have experienced worldwide growing popularity among people and organizations. Previously developed for military application, nowadays, companies and government entities make use of UAVs for a wide range of services, like localizing terrestrial objects~\cite{sorbelli2018range}, environmental monitoring and sensing~\cite{saffre2022monitoring}, search and rescue~\cite{barry2021information,careem2020rfeye}, public safety communications~\cite{geraci2022will}, delivery of small goods in urban areas~\cite{valencia2022factors,sorbelli2022scheduling}, among others. However, UAVs network infrastructure still faces a long way to achieve reliable basic~services, mainly due to mobility and positioning aspects.  For instance, UAVs may attain a speed of 460~km/h~\cite{han2009optimization} with high maneuverability in three-dimensional (3D) directions. While they get their global localization from Global Positioning System (GPS) signals or a known landmark~\cite{lima2021development}, obstacles in urban environments and other air vehicles in the same region may often make UAVs operations unfeasible.  Thus, they need to maintain a spatial awareness, i.e., to know the presence of others during flight. This knowledge allows them to choose a suitable trajectory in order to avoid collisions. In such conditions, a resilient and safe exchange of location is pivotal for the success of the services provided by the UAVs.

The safe flight of UAVs counts on various factors, like transported cargo volume and displacement speed~\cite{ulku2019sharing}, which directly affect device mobility. In this way, collision avoidance strategies allow UAVs to deal with undesirable events throughout their trajectory. Therefore, sharing UAV location is crucial for a coordinated flight, like the swarms of UAVs in aerial demonstrations~\cite{swarm2020}, in which many devices fly very close and must keep a safe distance among them~\cite{zhang2018adaptive} to perform air maneuvers. In this dynamic environment, the freshness of UAV location, i.e., how close to zero (current) is the age of such information from the last update~\cite{yates2021age} is essential to ensure UAVs' spatial awareness. Hence, sharing UAV locations resiliently in the face of device mobility and eventual connection loss improves Flying ad Hoc Networks' (FANET) operation.

The main existing studies on providing location-sharing services to UAVs focus on keeping them aware of others' locations throughout the operation. However, devices' mobility prevents them from performing this task safely. Some distributed solutions for multi UAVs operation employ tokens, i.e., messages circulating among devices, to carry their location~\cite{bekmezci2015location, ulku2016multi}. However, location sharing calls for proper token management to deal with network scalability~\cite{ulku2019optimization}. In the face of token collisions, using multiple channels improves UAV communication~\cite{ulku2019sharing}. Further, UAVs leverage intrinsic network services to broadcast their location embedded in the Service Set Identifier (SSID) field of beacon packets~\cite{minucci2019uav}. In contrast, centralized approaches employ a UAV with access to GPS signals to provide location to others~\cite{zhang2021multi, de2021uavouch}. In loss communications environments, adaptive methods boost location sharing ~\cite{zhang2018adaptive}. Lastly, clustering UAVs enhances network stability and accuracy to share location~\cite{bhandari2020mobility}. The literature discussed above enables UAVs to share their location. However, UAVs' mobility and autonomy hamper the usage of a central entity to monitor and control UAVs' positioning. Therefore, researchers pursue distributed and collaborative information-sharing approaches.

To address the aforementioned challenges, the age of UAVs' location is pivotal to their spatial awareness. Since their high mobility increases location updates, taking into account the freshness of location information improves UAVs' accurate and effective decision-making. The uncertainty of wireless channel states caused by UAV mobility leads to frequent network disconnections, so a timely location delivery by opportunistic approaches emerges as a viable way to deal with location delivery and updates. Further, given that UAVs need to know others' locations in order to fly safely, a location service must be resilient to overcome the impact of false locations in UAV operations. Therefore, our objective is to maximize neighbor discovery by considering their mobility behavior despite false location injected into the network. To achieve this goal, we have employed a joint optimization approach to discover and update close device information. In this work, we propose FlySafe, a resilient UAVs location sharing service based on information freshness and opportunistic deliveries to enable these vehicles to provide an application service. FlySafe leverages UAVs' active collaboration to timely share their location with other close devices. By joining opportunistic approaches like crowdsourcing and direct delivery, UAVs achieve spatial awareness by leveraging their mobility to deliver location updates. Hence, these vehicles achieve an optimized flight operation~\cite{wu2022crowdsourcing}, thus improving the communication efficiency of the whole system~\cite{lin2018sybil,karaliopoulos2019optimal}. Using mobile crowdsensing (MCS) approaches~\cite{karaliopoulos2019optimal,montori2023privacy} boosts UAVs' positioning sensing capabilities, thus improving their spatial awareness.

The remainder of this paper is organized as follows. Section~\ref{sec:related} gives an overview of related work. In Section~\ref{sec:problem}, we describe the system model and formulate the problem. Section~\ref{sec:propposal} presents the proposed solution. In Section~\ref{sec:evaluation}, simulation results and analysis are presented.
In Section~\ref{sec:conclusion}, we give conclusions and future work. In addition, the main notations are in Table~\ref{tab:notations}.

\section{Related Work} 
\label{sec:related}

In this section, we briefly summarize the studies aiming to share location in UAV networks and discuss the main shortcomings to help guide us in further improvements. In this way, it is possible to group state-of-the-art solutions into token circulation, broadcasting, and clustering. We refer to each group in the following subsections.

\subsection{Token Circulation-Based}

In~\cite{bekmezci2015location}, UAVs share location in FANETs by circulating~multiple tokens. Each one circulates by regions with closer devices to face network scalability issues. Upon receiving a token or overhearing exchanged messages, a UAV holds the neighbor's location and forwards the token with its location to others. In~\cite{ulku2016multi}, the seamless and accurate token reception relies on a bit of error control. The schema checks the bit error rate (BER) over data transmission, thus keeping a bit of control error for each package to resend in the face of any fault. Although both schemes allow location sharing in FANETs, there is a lack of information for token management. Further, UAV mobility compromises token exchange.

In~\cite{ulku2019sharing}, a schema addresses token collisions through a two-channel model, in which tokens circulate in one channel, whereas routing data flows through the other. The packet transmission occurs according to a threshold based on a BER control. While the second channel reduces token collisions, it delays routing message processing, which increases with the expansion of the number of tokens. Besides, this schema disregards the impact of UAV mobility on the exchanged messages. In~\cite{ulku2019optimization}, a master UAV manages tokens in a FANET, which act as vehicles and carry out routing planning. They visit all UAVs to share their location, so that strategy resembles the traveling salesman problem (TSP). As TSP is an NP-hard problem~\cite{lenstra1981complexity}, this approach handles token circulation as a combinatorial optimization problem. Although it optimizes token circulation, the master UAV becomes a single point of failure. Moreover, UAVs call for more processing resources to perform such functions.

\subsection{Supported by Broadcasting}

In~\cite{zhang2018adaptive}, UAVs share location to ensure swarm coordination, where a single relay allows to achieve a low payload and fast convergence in heterogeneous network environments. The approach employs an algorithm according to the network conditions: a lazy one to allow UAVs to broadcast data messages under request or an eager one to enable broadcasts whenever the requested data is available. Hence, the communication process is asynchronous and discretized, so the UAV with the strongest channel becomes a relay forwarding. Although the strategy improves data sharing, mainly in lossy environments, the network scalability increases the swarm convergence time for the desired information.

In~\cite{minucci2019uav}, UAVs share their location and speed embedded in the SSID of a Wi-Fi network. The broadcast of beacon packets enables UAV separation, like in the Automatic dependent surveillance-broadcast (ADS-B) technology for crewed aircraft. The receiver autonomously decides which channel to listen to, while the transmitter spans all the available channels. Thus, it takes more energy to re-transmit the same message on different channels. Such a strategy to share UAV locations reduces network overhead but requires firmware updates.

\subsection{Gathered by Clustering}

In~\cite{bhandari2020mobility}, cluster UAVs based on their location improves resource-constrained UAVs' performance and reliability in FANETs. Since message exchange commonly occurs inside clusters, such an approach reduces network overhead and latency. A UAV is selected as cluster head (CH) by centrality, proximity, randomness, mobility, and residual energy, while the others become members of the closest CH cluster. CHs communicate with sink nodes through routes created by a minimum spanning tree. The cluster maintenance takes place periodically according to UAVs' speed, distance to the CH, and distance from the cluster to the sink nodes. Despite the reduction in network overhead, UAVs' mobility frequently modifies network topology, thus compromising location exchanges.

In~\cite{zhang2021multi}, UAVs share their relative location to perform a coverage task collaboratively. A UAV group leader gets its position from GPS signals, while the others estimate their position from the leader's location by visual odometry. Then, UAVs share their local estimates through ultra-wideband communication. Based on the UAV's global position, the schema calculates the control inputs for every device in order to achieve optimal multi-UAV area coverage. Therefore, UAVs rely on leader's position to get local estimates, so the leader becomes a single point of failure. Further, UAVs require hardware resources to perceive a leader's position and compute their global position.

In~\cite{de2021uavouch}, validating UAVs' identity and location protects the network against malicious UAVs. A public-key-based authentication mechanism combined with movement plausibility checks for groups of UAVs operating on a battlefield close to an armored vehicle, which generates, encrypts, and sends session keys to each UAV by its public key. Vehicle position is a reference to UAV checks. Upon receiving a request from a new device to access the network, a UAV submits the decision to others in the group. Each UAV shares its location, speed, and acceleration with others to check the position of new UAVs. Thus, the position validation protocol relies on the relative positions of UAVs in the group and the position of the armored vehicle. The operation of UAVs with a land vehicle as a base station contributes to detecting malicious UAVs trying to access the network. However, the armored vehicle becomes a single point of failure.

Based on the literature above, the issue of sharing location in FANETs is a critical task for UAVs' safe operation. However, most existing research fails to consider the potential failures that can occur during location sharing, the coverage time limitations imposed by device mobility, and the security aspect. These studies rely on token circulation, which increases the delay on location updates and restrains network scalability; on broadcasting, which depends on firmware updates and increases network overhead; and on clustering, which requires frequent maintenance according to device mobility. To address these limitations, we argue that opportunistic approaches allow leveraging the freshness of UAV location, thus assisting in accurate and effective decision-making. Therefore, we propose a network model that takes into account UAV mobility, the freshness of UAV location, the age of location information, and security concerns. Then, we employ opportunistic approaches like crowdsourcing and direct delivery to improve UAV's location delivery in FANETs.

\section{Environment Scenario and Problem Formulation} \label{sec:problem}

In this section, we first analyze the environment scenario where UAV location sharing takes place. Then, we introduce the threat model devices are subject to during operation. Lastly, we present the problem formulation to maximize UAVs' knowledge about others.

\subsection{Environment scenario}

\begin{table}[!t]
	\centering
	\caption{Main Notations Used in This Paper}
        \relsize{-1.0}
        \begin{tabular}{m{2.2cm} p{10cm}}
		\toprule[1pt]
		\textbf{Notation} & \hspace{2.0cm}\textbf{Explanation} \\
		\midrule[0.5pt]
		$u$ & A UAV node \\
            $u_m$ & A UAV malicious node \\
            $Id_u$ & The identification of UAV $u$ \\
            $x^{t}_u,y^{t}_u,z^{t}_u$ & Coordinates of a UAV $u$ in time slot $t$ in a 3D space\\
            $L^{t}_u$ & The location of a UAV $u$ in time slot $t$ \\
		$\Ub$ & The set of deployed UAV \\
  		$\Ub^*$ & The set of malicious UAV  \\
            $\Wb_u$ & The set of neighbor nodes of a UAV $u$ \\
            $\Wb_{u}^*$ & The set of malicious neighbor nodes of a UAV $u$ \\
            $\Lb_u$ & The set of locations of a UAV node $u$ \\
            $\Tb$ & The set of time slots $t$ of a UAV operation \\
            $NL_u$ & The neighbor list of a UAV node $u$ \\
            $R$ & UAV nodes coverage radius \\
            $SL_u$ & The suspicious neighbor list of a UAV node $u$ \\
            $\lambda^{t}_u$ & Duration of time slots $t$ \\
            $\tau$ & Total mission completion time of a UAV \\
            $f$ & HM frequency \\
            $q_u$ & Quality flag of a neighbor node $u$ \\
            $d_u$ & Euclidean distance of a neighbor node $u$ \\
            $h_u$ & Hop flag of a neighbor node $u$ \\
            $a_u$ & Attitude flag of a neighbor node $u$ \\
            $s_u$ & State flag of a neighbor node $u$ \\
        \bottomrule[1pt]
	\end{tabular}
	\label{tab:notations}
\end{table}

\begin{figure}[!htb]
\centering
\includegraphics[width=0.75\textwidth]{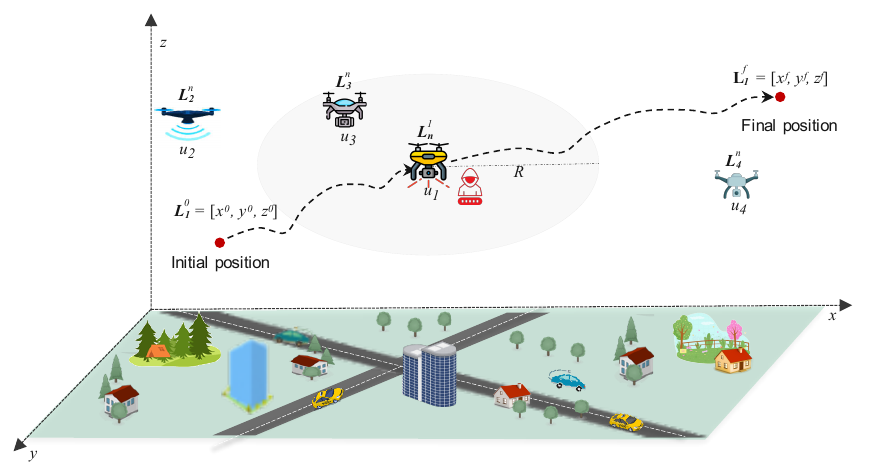}
\caption{UAV environment}
\label{fig:aerialScenario}
\end{figure}

We suppose an aerial environment, as shown in \mbox{Fig.~\ref{fig:aerialScenario}}, with several UAVs $u$ moving to provide an application service like monitoring~\cite{saffre2022monitoring}, and search and rescue~\cite{barry2021information,careem2020rfeye}, where $\Ub$ is the set of all UAVs available, $\Ub \triangleq \{1,2,...,U\}$, and $U$ is the total number of UAVs. The location of a UAV $u$ in a time slot $t$ is denoted by $L^{t}_u = \left[{x^{t}_u},{y^{t}_u},{z^{t}_u}\right]$, where ${x^{t}_u}$, ${y^{t}_u}$ and ${z^{t}_u}$ are $u$ coordinates in x-axis, y-axis and z-axis respectively. In this way, $L^{t}_u \in \Lb$, where $\Lb \triangleq \{1,2,...,L\}$, $t \in \Tb \triangleq \{1,2,...,T\}$ and $T$ is the total number of time slots. Taking into account that $L^{0}_1$ and $L^{f}_1$ are the initial and final locations of the UAV $u_1$, respectively, its trajectory can be approximated by $\{L^{0}_1,L^{t}_1,L^{2t}_1,...,L^{f-t}_1,L^{f}_1\}$.

The duration of a time slot $t$ is $\lambda^{t}$, so that the total mission completion time of a UAV $u$ can be given by $\tau = \sum_{n=1}^{T} \lambda^{t}_u$. Therefore, $\lambda^{t}_u$ must be determined sufficiently short, i.e., $\lambda^{min} \leq \lambda^{t}_u \leq \lambda^{max}$, such that changes in UAV location within each $t$ is significant for other UAVs to recognize it. We also assume all UAVs operating with time slots of the same duration $\lambda^{t}_u$ $\forall u \in \Ub$. Hence, we represent the trajectory of a UAV $u$ by $T$ waypoints in 3D coordinates as $L^{t}_u$ $\forall t \in \Tb$, together with $\lambda^{t}_u$ $\forall t \in \Tb$.

Moreover, we take that each UAV is aware of its location information by using GPS within tolerable error, which means that UAV $u_i$ has a vector of 3D position $\textbf{L}_{i}^t\in\mathbb{R}^{3\times1}$ in time slot $t$. Supposing that the UAV communication channels takes place by line of sight (LoS) links, free of obstacles, we disregard any path loss model and interference for link between two UAVs, like $u_i$ and $u_j$. The distance between them $d_{i,j}$ is the Euclidean distance calculated by (\ref{eq:euclidean}), where $p$ corresponds to the coordinates in x-axis, y-axis and z-axis.

\begin{equation}\label{eq:euclidean}
	d_{i,j}=\sqrt{\sum\nolimits_{k=1}^3\left|\,p_{i,k}-p_{j,k}\right|^2}
\end{equation}

During $u_1$ trajectory, as shown in Fig.~\ref{fig:aerialScenario}, it sends signals according to a wireless technology with a transmission range $R$. Thus, this UAV establishes network connections with others $u_n$ inside its coverage area in a time slot $t$, such that $d_{i,n}^t \leq R$. The neighborhood of $u_1$, $\Wb_1^{t}$, encompasses all $u_n$, where \mbox{$\Wb_i \triangleq \{1,2,...,W\}$}. $W$ is the total number of $u_n$ and $W \leq U-1$. Therefore, $u_1$ achieves spatial awareness whenever it has knowledge about its neighborhood $\Wb_1^{t}$, i.e., it recognizes all nodes in its coverage area. To keep spatial awareness, $u_1$ must update $\Wb_1$ periodically according to time slots $t$.

\subsection{Threat model}

Whenever a malicious UAV (MalUAV) carries out false data injection (FDI) attack, it shares a false location (FL) with its neighbors to damage their spatial awareness. Thus, it jeopardizes the location service and makes UAV operations~unsafe. We consider the behavior of a UAV performing a FDI attack during its flight operation. We denote the set of all MalUAVs $m$ available at time slot $t$ as $\Ua^t$, such that $\Ua^t \subset \Ub$. $\Wa_{i}^t$ means the malicious neighborhood of a UAV $u_i$ at $t$. It encompasses all MalUAVs $m$ inside $u_i$ coverage area at $t$, where $m \in \Wa_{i}^t$ and $m \in \Ua_{i}^t$. Therefore, as a MalUAV $m$ sends a false location $L^{t}_m$, $L^{t}_m \notin \Lb$.

We consider the scenario depicted in \mbox{Fig.~\ref{fig:aerialScenario}}, where UAV $u_1$ travels a trajectory from $L^{0}_1$ and $L^{f}_1$ approximated by $\{L^{0}_1,L^{t}_1,L^{2t}_1,...,L^{f-t}_1,L^{f}_1\}$. Whether $u_1$ performs FDI, it shares FL with others throughout the trajectory to deceive them about its true position. For instance, whenever $u_3$ receives an FL, it registers $u_1$ in its malicious neighborhood $\Wa_{3}$. FL then compromises $u_3$ spatial awareness, seriously threatening both UAV operations. In such conditions, it is pivotal to a safe operation to identify a UAV sharing FL and segregate it from the network. Therefore, the location service becomes resilient to FL and improves UAVs' spatial awareness.

\subsection{Problem formulation}

Let \textbf{U}~$=\{U^{t}_u, t \in \Tb\}$, \textbf{L}~$=\{L^{t}_u, t \in \Tb\}$, and \mbox{T~$=\{\lambda^{t}_u, t \in \Tb\}$}, we model the optimization problem as

\begin{subequations}{\label{optimal}}
\begin{align}
    \Pb:  \max_{\mathbf{L}, \mathbf{T}, \mathbf{U}}&\text{~} W=\sum_{u \in \Wb} u,\tag{2} \label{eq:problem}\\
       \textrm{s.t.}\; 
       & d_u \leq R \text{  } \forall u \in \Wb,\tag{2.1} \label{eq:const21} \\
       & \Wb \subset (\Ub - \Ua) \text{  } \forall u \in \Wb, \tag{2.2} \label{eq:const22}\\
       & \sum_{u \in (\Wb - \Wa)} u < \sum_{u \in (\Ub - \Ua)} u \text{  } \forall u \in \Wb, \tag{2.3} \label{eq:const23}\\
       & \lambda^{min} \leq \, \lambda^{t}_u \leq \lambda^{max} \text{  } \forall t \in \Tb, \tag{2.4} \label{eq:const24}\\
       & L^{t}_u \neq  \, L^{t+1}_u \text{  } \forall L_u \in \Lb \text{,  } \forall u \in \Ub \text{~and~} \forall t \in \Tb.\tag{2.5}\label{eq:const25}
\end{align}
\end{subequations}

\noindent
where we aim to maximize the number of neighbors $W$ of a UAV; (\ref{eq:const21}) is the constraint to ensure that a neighbor is inside the coverage area of UAV $u$, i.e., $d_u \leq R$; (\ref{eq:const22}) is the constraint to ensure that the neighborhood $\Wb$ is a subset of the UAVs $u$ in the aerial environment, $u \in \Ub$, disregarding MalUAVs $m \in \Ua$; (\ref{eq:const23}) is the constraint to ensure that the size of the neighborhood $\Wb$ is less than the size of the set of UAVs, $\Ub$, disregarding MalUAVs $m \in \Ua$; (\ref{eq:const24}) is the constraint to ensure that the duration of time slot $t$, $\lambda^{t}_u$, allows UAVs to recognize location changes; (\ref{eq:const25}) is the constraint to ensure that all UAVs are moving and share true location. One can see that (\ref{eq:problem}) is a convex problem, as the challenge is discovering the nearest UAVs and updating this information. However, the target function of $\Pb$ is somewhat untrackable, mainly because the UAVs displacement model is unpredictable, and they do not know each other beforehand. To partially solve $\Pb$ using traditional device discovery protocols will cost enormous overhead to the network, often making it impossible to be online deployed. Due to this reason, next, we will propose a solution to address problem $\Pb$.

\section{FlySafe - A Resilient UAVs Location Sharing Service}
\label{sec:propposal}

This section details FlySafe, a resilient UAVs location sharing service based on information freshness and opportunistic deliveries to make UAVs aware of others in their coverage area.  In this way, UAVs conduct a proper flight coordination, thus preventing collisions. Firstly, we present FlySafe setting model. Next, we detail the communication model. Then, we describe FlySafe and its architecture.

\subsection{FlySafe setting model}

We consider that FlySafe runs between network and application layers in order to enable UAVs to share their location with others. Hence, they make distributed and autonomous decisions to report their presence and mobility behavior. UAVs operate in two phases: \textit{Neighbor discovery} and \textit{Neighbor maintenance}. A network discovery protocol combined with opportunistic approaches like crowdsourcing and direct delivery perform for UAVs throughout both phases. It allows UAVs to identify their neighborhood and keep it updated. In this regard, they dynamically adjust their sensing scheme based on their mobility behavior to obtain accurate location information from others with a tolerable network overhead. Therefore, the operation of the joined approaches enables resilient location information sharing, leading UAVs to achieve spatial awareness. 

\begin{figure}[!htb]
\centering
\includegraphics[width=0.9\textwidth]{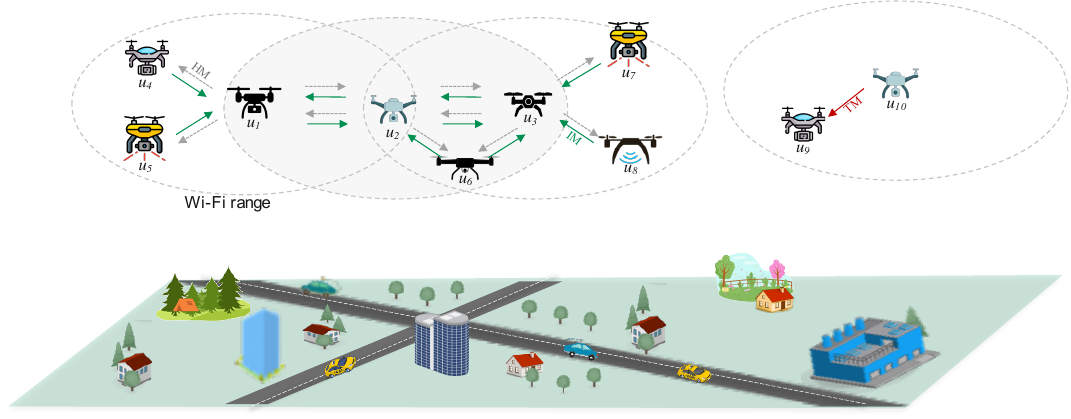}
\caption{Environment overview}
\label{fig:overview}
\end{figure}

In our operation setting, as shown in Fig.~\ref{fig:overview}, FlySafe runs over a set of UAVs interconnected by a wireless communication network, denoted by \mbox{$\Ub = \{u_1, u_2, u_3, ..., u_j\}$}, where $u_j \in \Ub$. We take all UAVs have communication and processing resources to disseminate data, and each one owns a unique address ($Id$) to identify it in all time slots $t$. Further,  they get their location from GPS, and the wireless channel for connecting each other operates by a LoS wireless link.

For instance, the neighborhood information of a UAV $u$ in a slot $t$, $\Wb_u^{t}$,
encompasses information of others, each one represented by a set of distinct tuples $\langle Id, t, L^{t}, q^{t}, h^{t}, a^{t}, d^{t}, s^{t}\rangle$, such that

\vspace{0.25cm}

\noindent
\scalebox{0.8}{
$\Wb_u^{t} =\{\langle Id_1, t, L^{t}_1,q^{t}_1, h^{t}_1, a^{t}_1, d^{t}_1, s^{t}_1\rangle, \langle Id^{t}_2, t, L^{t}_2, q^{t}_2, h^{t}_2, a^{t}_2, d^{t}_2, s^{t}_2 \rangle, ...,\langle Id_j, t, L^{t}_j, q^{t}_j, h^{t}_j, a^{t}_j, d^{t}_j, s^{t}_j\rangle\}.$}

\vspace{0.25cm}

\noindent
We denote the location of each UAV $u$, $L^{t}_u$, as a tuple $\langle x^{t}_u, y^{t}_u, z^{t}_u\rangle$, meaning its position in a 3D space. The \textit{quality flag} $q_u$ value shows how fresh is the neighbor location received and kept in the neighbor list (NL). It points out the update frequency of a neighbor location being $q_u$ an integer value between 0 and 3, so that value 3 means the neighbor has updated its location in the last time slot $t$, whereas the value 0 implies no location updates in the previous $3t$. The \textit{hop flag} $h_u$ value holds whether the neighbor is within a coverage area of a UAV, $h_u = 1$, or beyond it, $h_u > 1$. Hence, it helps to extend UAV's spatial awareness by enabling knowledge about others outside the coverage area. The \textit{attitude flag} $a_u$ value represents a neighbor mobility behavior relative to a UAV, i.e., inbound, $1$; outbound, $2$; or a static position, $0$. The \textit{distance} $d_u$ equals the Euclidean distance between a UAV and one~neighbor. The \textit{state} $s_u$  value means whether the neighbor UAV is ordinary, 0, or malicious, 1. By simplicity, we assume that a turned-off or a stationary UAV on the ground is out of the network. Further, any UAV can act maliciously and make an FDI attack by sharing false location in order to jeopardize the FlySafe~operation.

\begin{figure}[htb]
\centering
\includegraphics[width=0.75\textwidth]{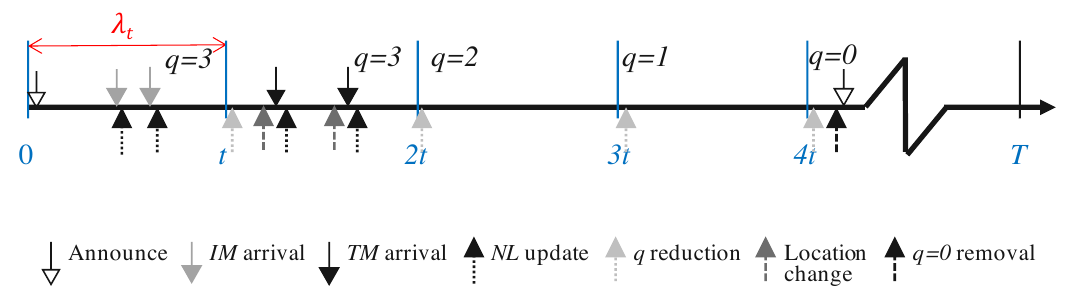}
\caption{Evolution of network events}
\label{fig:events}
\end{figure}

As all UAVs move in time and space, they can simultaneously be a neighbor of one or more UAVs according to their location, and they can also have several neighbors. Each UAV starts by discovering its neighborhood and keep it updated. Hence, it can establish \textit{ad hoc} network connections with others to exchange messages. However, a neighbor moving outbound from a UAV coverage area eventually interrupts established connections. As shown in Fig.~\ref{fig:overview}, UAV $u_1$ can establish \textit{ad hoc} networks to connect with various neighbors, like $u_2, u_4, u_5$, whereas $u_2$ also belongs to $u_6$ neighborhood, $\Wb_6 = \{u_2,u_3\}$. Fig.~\ref{fig:events} presents a timeline of network events during neighbor discovery and maintenance phases, which we discuss below. 

\vspace{0.1cm}
\subsubsection{Neighbor discovery (ND)}

Neighbor discovery protocols usually identify devices connected to the network by frequently exchanging hello messages~\cite{lakew2020routing}. Although they suits many settings, the UAVs~flight behavior often compromises messages reception. Such condition jeopardizes the identification of others. The dynamic environment calls for more exchanged messages among UAVs during neighbor discovery and neighborhood maintenance phases. In this sense, FlySafe focuses on the scenario 
seen~in~Fig.~\ref{fig:overview}.

FlySafe makes usage of  two messages to discover neighbor UAVs, called {\it Hello message (HM)} and {\it Identification message (IM)}, and on the quality flag $q$ to compel NL updates. Whenever a UAV $u$ starts operating, as shown in Fig.~\ref{fig:events}, it announces its presence by broadcasting an HM  $\langle Id, L_u, NL_u \rangle$, that hold its identification $Id_u$, location $L_u$, and $NL_u$, respectively. Moreover, during UAV operation, a new announcement takes place whenever $NL_u$ becomes empty, $|\Wb_u| = 0$, or when it holds only one-hop away neighbors, $h > 1$. Hence, FlySafe reduces message overhead to perform neighbor discovery. Given that standard off-the-shelf UAVs commonly move lower than crewed aircraft, we set up $\lambda^{t}_u$ to 1.5 seconds. Such period equals three times ADS-B systems message report time ~\cite{icao2018} for crewed aircraft. Further, this time slot duration allows UAVs to~send HM with a frequency $f = 1/\lambda^{t}_u$ whenever it announces its presence.

The \textit{Identification message} (IM) is a unicast response to a received HM. Upon the reception of
an HM, a UAV updates its $NL$ with the announcer information ($Id, \, L = location,$ $q = 3,  \,  h = 1,  \, a = 0,  \, d = calculated \, distance,  \, s = 0$). Next, it answers with an IM $\langle Id, L_u, NL_u \rangle$, that hold its identification $Id_u$, location $L_u$, and $NL_u$, respectively. In the same way, whenever an announcer UAV receives an IM, it registers in 
its $NL$ the neighbor information ($Id, \, L = location, \, q = 3,  \,  h = 1,  \, a = 0,  \, d = calculated \, distance,  \, s = 0$). Therefore, both messages -- HM and IM -- enable simultaneous neighbor discovery, i.e., the announcer UAV discover others, whereas the neighbor perceives the presence of the announcer. For instance, as shown in Fig.~\ref{fig:events}, neighbor discovery phase occurs in the first time slot when a UAV starts announcing its presence. Then, it updates its $NL$ after receiving IM from UAVs in its coverage area.

\vspace{0.1cm}
\subsubsection{Neighbor maintenance (NM)}

UAVs' spatial awareness counts on up-to-date neighbor location. Hence, we adopt UAVs mobility as a trigger to update others about location changes and the \textit{quality flag} $q$ to compel NL updates. Therefore, whenever a UAV changes its location, it sends a \textit{Trap message} (TM) $\langle Id, L_u, NL_u \rangle$ to one-hop away neighbors in its NL. Such a unicast message carries UAV identification $Id_u$, updated location $L_u$, and $NL_u$, respectively. Upon receiving a TM, a UAV updates its NL with the neighbor's new location, other information ($q = 3,  \,  h = 1,  \, a = 0,  \, d = calculated \, distance,  \, s = 0$), and with the UAVs information in neighbors NL.

Every UAV starts NM at the beginning of each time slot $t$, as depicted in Fig.~\ref{fig:events}. Initially, it decreases by one the \textit{quality flag} $q$ value of all neighbors within its NL. Next, it removes from NL the neighbors with $q = 0$. Therefore, such successive $q$ reduction ensures that a neighbor will remain in a NL for at least $3t$ after being discovered. In the case of an empty NL, the UAV performs ND. For instance, as shown in Fig.~\ref{fig:events}, NM takes place from time slot $t$ with the $q$ reduction in all slots. Due to the absence of neighbor updates from slot $2t$ to $4t$, the $q$ flag attains $0$ for all neighbors in slot $4t$. Consequently, the UAV removes all neighbors from the NL, which becomes empty, and perform ND in slot $4t$.

\vspace{0.1cm}
\subsubsection{Network connections evolution}

\begin{figure}[!t]
\centering
\includegraphics[width=0.75\textwidth]{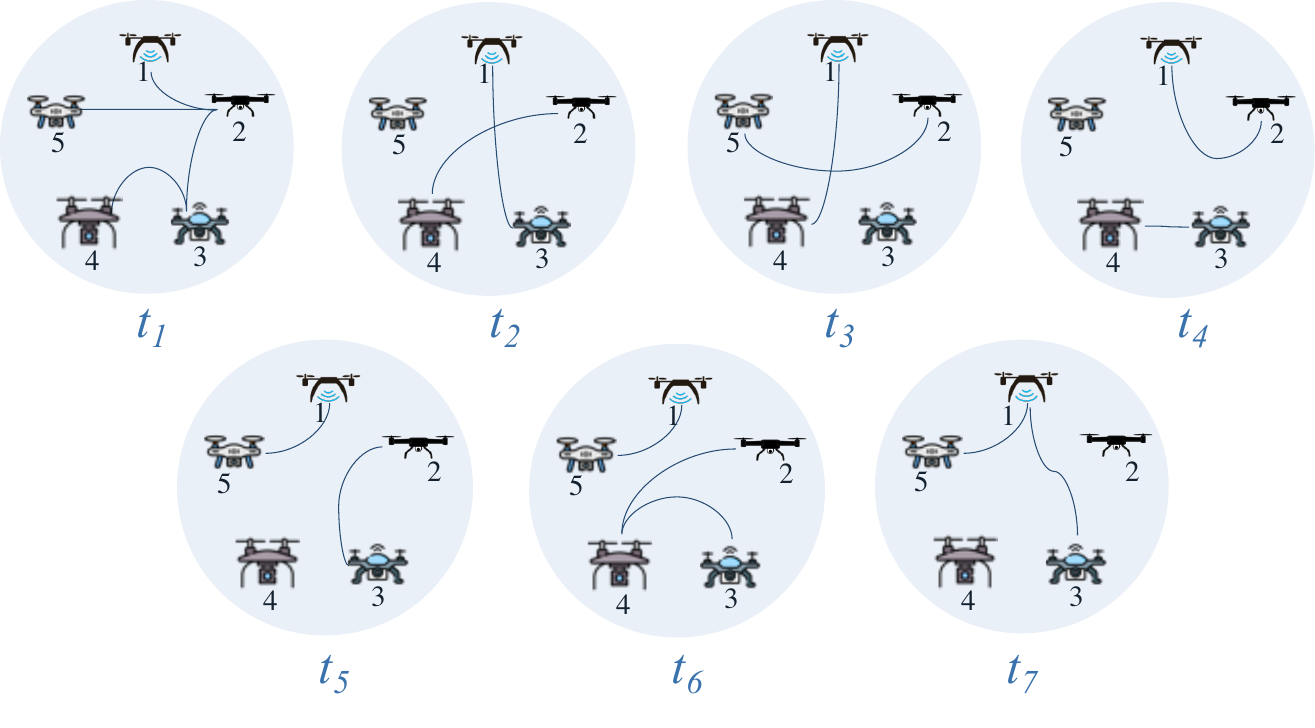}
\caption{Evolution of UAV network connections}
\label{fig:graphs}
\end{figure}

Taking into account that UAVs establish network connections under a dynamic behavior in spatial and temporal dimensions, we take their interactions through dynamic graphs or temporal graphs. Thus, we denote a static network by a graph $G (V,E)$, where a set of vertices $V$ corresponds to UAVs and a set of edges $E:V \times V \Rightarrow \mathbb{R}$ represents UAVs connections. Given that a UAV neighborhood changes over time due to devices' mobility, we represent it through several graphs $G_1, G_2, ..., G_T$, each according to a time slot $t \in \Tb$ of network operation. For instance, Fig.~\ref{fig:graphs} represents the evolution of network connections in a network with five UAVs, where $V = \{1,2,3,4,5\}$. As the edges $E$ change according to network connections, we have a graph with distinct links at each time slot $t$. In time $t_1$, $E(G_1)=\{(1,2),(2,3),(2,5),(3,4)\}$. However, in time slot $t=7$, UAVs establish other connections and form a new graph, $G_7$, such that $E(G_7)=\{(1,3),(1,5)\}$.

We represent a temporal graph $G$ by $G(\Tb,V,N,W,E)$, where $\Tb$ is a finite set of discrete or continuous time slots, $V$ is a finite set of vertices, $N$ is a set of temporary UAVs (vertices), such that $N \subset \Tb \times V$, and $E$ is the set of links $E \subset V \times V$, so that $(t,uv) \in E$, implying $(t,u) \in E$ and $(t,v) \in E$. Thus, the neighborhood $W$ of a vertex $v$ in $G$ corresponds to $W(v)$, where $v \in V$ and $u \in V$, \mbox{$W(v)=\{u, uv \in E\}$}. The degree $d$ of a vertex $v$, $d(v)$, means the number of UAVs close to the vertex $v$ at a given moment, i.e, the neighborhood size.

\subsection{Communication model}

We suppose a connectivity model so that UAVs start functioning unaware of others within their coverage area. Further, they have equal communication coverage and operate in the same transmission band. All UAVs can exchange messages whenever they are within the same coverage area. Additionally, packet losses can take place due to noise, UAV mobility, or other failures during device operation.

\begin{figure}[!htb]
\centering
\includegraphics[width=0.75\textwidth]{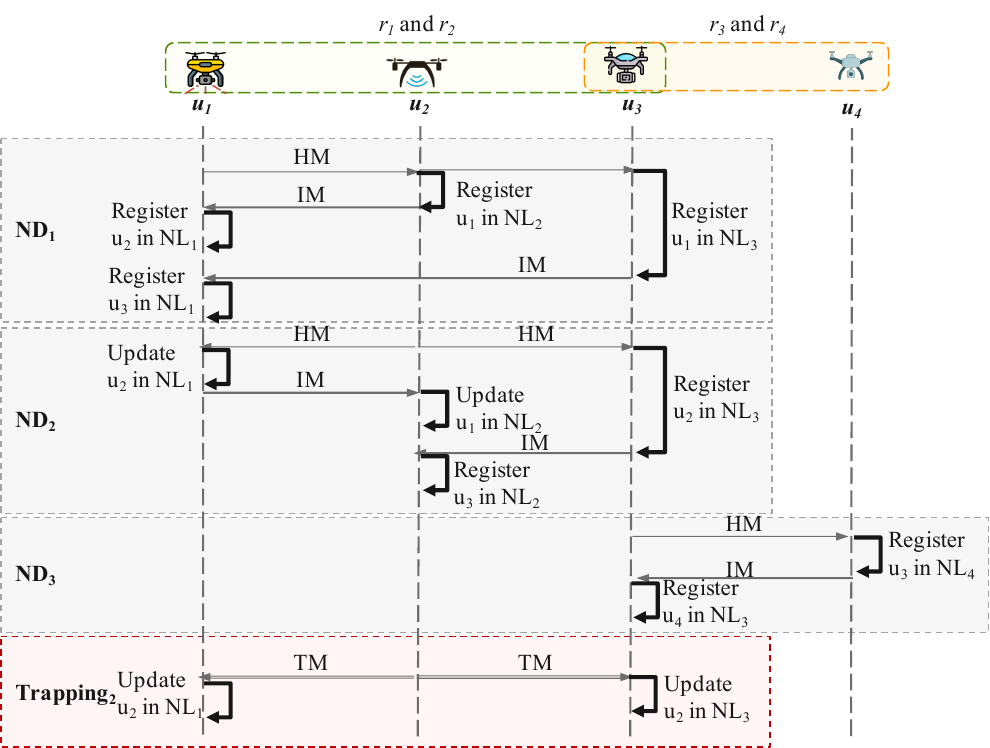}
\caption{Communication flow between UAVs}
\label{fig:commflow}
\end{figure}

The communication flow between UAVs occurs~as depicted in Fig.~\ref{fig:commflow}. UAVs $u_2$ and $u_3$ are in the $u_1$ coverage area $r_1$, while $u_1$ and $u_3$ are in the $u_2$ coverage area~$r_2$. UAV $u_1$ starts neighbor discovery (ND$_1$) by broadcasting an HM. Upon receiving such a message, $u_2$ registers $Id_1$ and $L_1$ in its NL. Next, it sends an IM to $u_1$, which stores $u_2$ information in its NL$_1$. UAV $u_3$ also receives the HM from $u_1$, registers it in its NL$_3$, and sends an IM to $u_1$. UAV $u_1$ saves $u_3$ in its NL$_1$. Then, $u_2$ and $u_3$ perform ND$_2$ and ND$_3$, respectively, to update their NL.

Whenever a UAV moves, it must update its neighbors with the new location in order to improve their spatial awareness. As illustrated in Fig.~\ref{fig:commflow}, once UAV $u_2$ detects a location change during its operation, it sends a TM to UAVs in NL$_2$ -- $u_1$ and $u_3$. Therefore, upon receiving the TM, $u_1$ and $u_3$ update $u_2$ location in their respective NL.

\subsection{System overview}

Spatial awareness means UAVs have knowledge about others flying around them. Hence, they fly safely and make proper decisions to provide a suitable application service like monitoring~\cite{saffre2022monitoring}, and search and rescue~\cite{barry2021information,careem2020rfeye}. In this regard, as depicted in Fig.~\ref{fig:overview}, UAVs interact to discover their neighborhood and keep it updated. In this scenario, for instance, UAV $u_2$ starts ND announcing its presence by broadcasting an HM. Upon receiving the HM, each UAV in $u_2$ coverage area -- $u_1$, $u_3$, and $u_6$ -- registers $u_2$ in its NL and answers with an IM $\langle Id, L, NL \rangle$, that hold its identification $Id$, location $L$, and $NL$, respectively.

Upon receiving an IM, each UAV retrieves the source UAV location to store this neighbor in its NL. Further, this UAV also executes this process to UAVs inside the neighbor NL. For instance, as depicted in Fig.~\ref{fig:overview}, $u_2$ receives an IM from $u_1$ and inserts $u_1$ in its NL. Then, it examines $u_1$ NL, which contains three UAVs -- $u_2$, $u_4$, and $u_5$ --, and inserts $u_4$ and $u_5$ in its NL as second hop neighbors. As $u_2$ also receives IM from $u_3$ and $u_6$, it performs the same way as $u_1$. After $u_2$ finishes ND, its NL contains seven UAVs: three first-hop neighbors -- $u_1$, $u_3$, and $u_6$ -- and four second-hop neighbors -- $u_4$, $u_5$, $u_7$, and $u_8$.

Considering $u_{9}$ and $u_{10}$ have already performed ND, thus being aware of others in their coverage area. From now on, they will share with their neighbors any location change during their operation. For instance, as illustrated in Fig.~\ref{fig:overview}, whenever $u_{10}$ moves to another location, it sends a TM to its neighbors. Upon receiving such a message, $u_{9}$ updates its NL with $u_{10}$ new location and neighbor information.

\subsection{Architecture} 
\label{sec:architecture}

\begin{figure*}[!htb]
\centering
\includegraphics[width=0.99\textwidth]{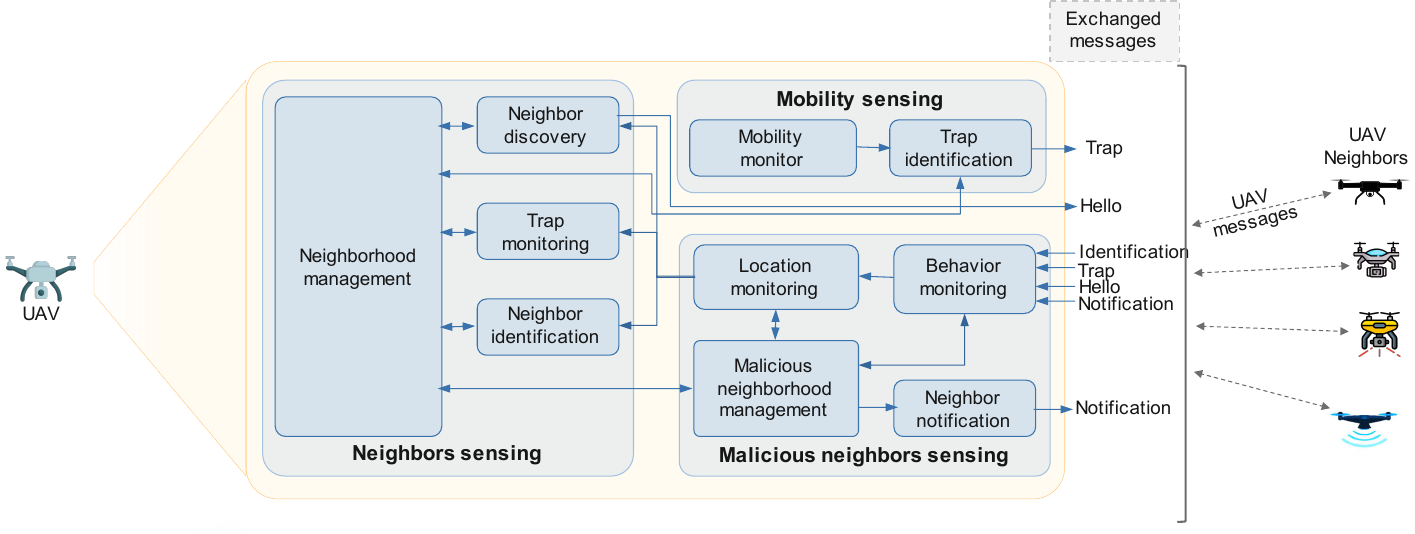}
\caption{System architecture}
\label{fig:architecture}
\end{figure*}

The FlySafe architecture comprises three modules installed on a UAV, as shown in Fig.~\ref{fig:architecture}. The \textit{Neighbors sensing} module performs the neighbor discovery and keeps the UAV updated about the existence and location of others within its coverage area. The \textit{Mobility sensing} module monitors the device mobility behavior and alerts its neighbors about any location change. The \textit{Malicious neighbors sensing} module monitors received messages to identify and keep the UAV updated about MalUAVs within its coverage area injecting false location.

\textit{1) Neighbors sensing module (NS)}: It provides UAV spatial awareness by identifying devices within its coverage area. In this sense, NS components determine these devices and hold their $Id$ and $L$ in an NL. NS updates neighbors' information by exchanging messages—HM, IM, and TM—with others moving in space and time within the UAV coverage area.

FlySafe performs ND as described in Algorithm~\ref{alg:nd}. In this way, all UAVs start operating and create their NL to keep neighbor information (\textit{l}.1). Then, they announce their presence by broadcasting an HM (\textit{l}.2). Whenever a UAV \textit{u} receives an HM, it responds to the announcer with an IM. Once \textit{u} receives an IM from a UAV \textit{n}, in the case of a new neighbor, it puts \textit{n} in its \textit{NL}. Otherwise, it updates \textit{n} location in NL (\textit{l}.4-9). Further, whether \textit{n} has neighbors, UAV \textit{u} updates its NL with \textit{n} neighbors ($n_n$) information (\textit{l}.10-18).

\begin{algorithm}[!htb]
{
    \DontPrintSemicolon
    \SetAlgoLined
    \footnotesize
    \KwData{\textit{u} (UAV device), \textit{n} (UAV neighbor device),
    ND (Neighbors Discovery), NL (Neighbors List), L (Location Information)}
    \textit{NL(u)} $\leftarrow 0$ \\
    \textit{Announce-u(L(u), NL(u))} \tcp*{\footnotesize{HM broadcasted}}
    \While(\hspace{6.5cm}\tcp*[h]{\footnotesize{UAV operating}}){\textit{u} = ON}{ 
        \textit{u}$\leftarrow$ \textit{L(n) + NL(n)} \tcp*{\footnotesize{IM received}}
        \uIf{\textit{n} $\notin$ \textit{NL(u)}}{
    	\textit{NL(u)} $\leftarrow$ \textit{n}
        }
        \Else{
    	\textit{NL(u)} $\leftarrow$ \textit{L(n)}
        }
        \If{\textit{NL(n)} $\neq \emptyset$}{
    	\For{$\forall$ \textit{n}$_n \in$ \textit{NL(n)}}{
    		\uIf{\textit{n}$_n \in$ \textit{NL(u)}}{
    			\textit{NL(u)} $\leftarrow$ \textit{L(n$_n$)}
    		}
    		\Else{
    			\textit{NL(u)} $\leftarrow$ \textit{n$_n$}
    		}
    	}
        }
}
\caption{Neighbors discovery management}
\label{alg:nd}
}
\end{algorithm}

\textit{2) Mobility sensing module (MS)}:  It manages the decision-making on UAVs' mobility, as described in Algorithm~\ref{alg:ms}. After the ND phase, the MS module works to keep updated neighbor information. During operation, whenever a UAV \textit{u} senses any location change (\textit{l}.1-2), case it has known neighbors, it decreases by one their quality flag $q$. Next, it removes from NL neighbors with $q = 0$ (\textit{l}.3-5). In the case of an empty NL or with only neighbors beyond UAV \textit{u} coverage area (\textit{l}.6-7), it makes ND or sends a TM to all neighbors in NL otherwise (\textit{l}.8-11). Whenever UAV \textit{u} is unaware of neighbor nodes, it performs ND (\textit{l}.11-12).

\begin{algorithm}[!htb]
{
    \DontPrintSemicolon
    \SetAlgoLined
    \footnotesize
    \KwData{\textit{u} (UAV device), \textit{n} (UAV neighbor device),
    ND (Neighbor Discovery), NL (Neighbors List), L (UAV location)}
    \While(\hspace{6.4cm}\tcp*[h]{UAV operating}){\textit{u} = ON}{
        \If(\hspace{3.6cm}\tcp*[h]{UAV moving}){\textit{Location\_change(u) $=$ True }}{
            \uIf{\textit{NL(u) $\neq \emptyset$ }}{
                \textit{Decrease $q$ by 1 $\forall$ n $\in$ NL(u)} \\
                \textit{Remove n with $q = 0$ from NL(u)} \\
                \uIf{(\textit{NL(u) $= \emptyset$}) || (h $> 1 \; \forall$ n $\in$ NL(u))}{
                    \textit{Perform ND} \\
                }
                \Else{
                    \textit{Send (L(u), NL(u)) $\forall$ n $\in$ NL(u)} \tcp*{TM sent}
                }
            }
            \Else{
                \textit{Perform ND} \\
                }
        }
    }
    \caption{Mobility sensing management}
    \label{alg:ms}
}
\end{algorithm}

\textit{3) Malicious neighbors sensing module (MM)}: It monitors the received messages to determine false locations sent by neighbor UAVs. As described in Algorithm~\ref{alg:fdi}, all UAVs start operating and create their Suspect List (SL) to hold information about MalUAVs (\textit{l}.1). Whenever a UAV \textit{u} receives a message, its provenance is first assessed to avoid processing messages from MalUAVs already blocked by the system (\textit{l}.3-5). In the case of a false location received from an honest neighbor \textit{n}, UAV \textit{u} turns it suspicious and puts it in SL. Next, UAV \textit{u} notifies all its neighbors about \textit{n} and processes the received message (\textit{l}.7-12). When the received message comes from a known MalUAV \textit{n} (\textit{l}.13), \textit{u} increases this \textit{n} recurrence by 1. Whether \textit{n} achieves a threshold of 3 false locations sent in a row (\textit{l}.14-15), \textit{u} blocks \textit{n} in the SL and removes it from NL. Next, it notifies others about the blocked UAV and drops the received message~(\textit{l}.16-18).

\begin{algorithm}[!htb]
{
    \DontPrintSemicolon
    \SetAlgoLined
    \footnotesize
    \KwData{\textit{u} (UAV device), \textit{n} (UAV neighbor device),
    NL (Neighbors List), L (Location), SL (Suspect List),
    \textit{r} (UAV recurrence), \textit{s} (UAV state)}
    \textit{SL(u) $\leftarrow 0$ } \\
    \While{\textit{u} = ON}{
        \textit{u $\leftarrow$ L(n) + NL(n)} \\
        \uIf{\textit{IsBlocked(n)} $=$ True}{
            \textit{Drop (L(n) + NL(n))} \tcp*{Drop messages from blocked UAVs} 
        }
        \Else{
            \uIf{\textit{L(n) $=$ False }}{
                \uIf{IsHonest(n) $=$ True}{ 
                    \textit{Set $s_n$ to 1 in NL(u)} \\
                    \textit{Insert $n$ in SL(u)} \\
                    \textit{Send $s_n$ $\forall$ n $\in$ NL(u)}  \tcp*{Notify on suspect UAV}
                    \textit{Process (L(n) + NL(n))} \\
                }
                \Else{
                    \textit{Increase $r_n$ by 1 in SL(u)} \\
                    \If{$r_n =$ 3}{
                        \textit{Set $s_n$ to 1 in SL(u)} \\
                        \textit{Remove \textit{n} from NL(u)} \\
                        \textit{Send $s_n$ $\forall$ n $\in$ NL(u)} \tcp*{Notify on blocked UAV}
                        \textit{Drop (L(n) + NL(n))} \\
                    }
                }
            }
            \Else{
                \If{\textit{IsSupect(n)} $=$ True}{
                    \textit{Decrease $r_n$ by 1 in SL(u)} \\
                    \If{$r_n =$ 0}{
                        \textit{Set $s_n$ to 0 in NL(u)} \\
                        \textit{Remove n from SL(u)} \\
                        \textit{Send $s_n$ $\forall$ n $\in$ NL(u)} \tcp*{Notify on honest UAV}
                    }
                }
                \textit{Process (L(n) + NL(n))} \\
            }
        }
    }
    \caption{FDI attacks management}
    \label{alg:fdi}
}
\end{algorithm}

Whenever a UAV \textit{n} sends a false location during a regular operation due to a GPS failure, for instance, such a condition makes it suspicious to its neighbors (\textit{l}.23). However, case \textit{n} sends a true location in a row, the receiver UAV decreases its recurrence by 1 (\textit{l}.24). When \textit{n} recurrence reaches 0, it means \textit{n} became honest. Hence, the receiver UAV \textit{u} changes the \textit{n} state to honest and removes it from SL. Then, the \textit{u} notifies others about \textit{n} (\textit{l}.25-28) and processes the received message~(\textit{l}.31).

\section{Evaluation and Analysis} \label{sec:evaluation}

This section presents the resilience and performance evaluation carried out by simulation in order to analyze the behavior of the FlySafe service. First, we detail the scenario and configurations applied in the simulations. Next, we present the metrics employed to evaluate FlySafe against FDI attacks. Lastly, we show and discuss the achieved results.

\subsection{Simulation Scenario}

\begin{figure}[!htb]
\centering
\includegraphics[width=0.79\textwidth]{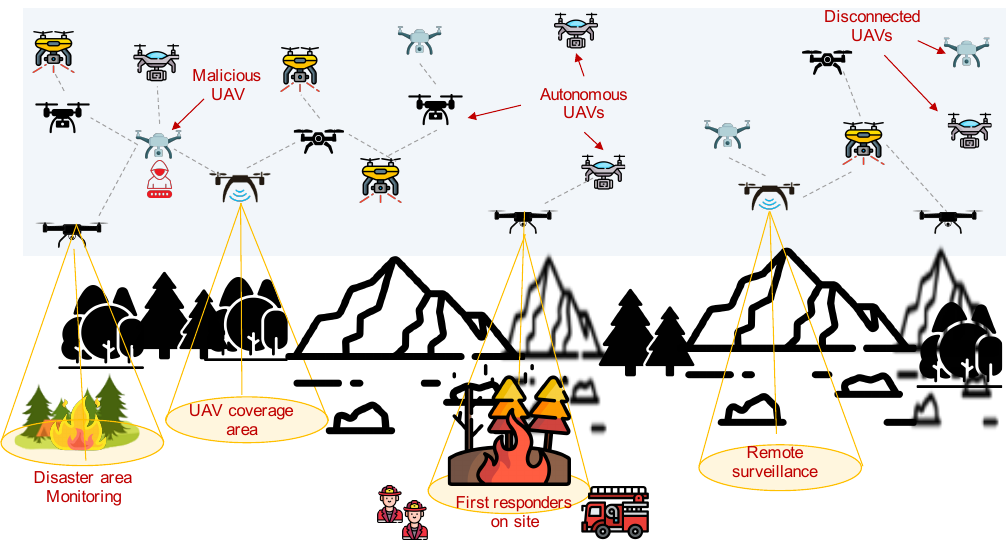}
\caption{Autonomous UAV network performing monitoring, search and surveillance in inaccessible disaster areas}
\label{fig:3Dscenario}
\end{figure}

We have employed the NS-3 network simulator~\cite{NS3simulator2022} to implement FlySafe and assess its performance and resilience against FDI attacks. We apply the simulation parameters presented in Table~\ref{tab:simsettings}. As depicted in Fig.~\ref{fig:3Dscenario}, we took a scenario in which a swarm of 40 UAVs monitors and surveys an inaccessible region with an area of \mbox{1.5 $\times$ 1.5~km}, where the fixed communication infrastructure is unavailable. Given the existing lack of standardization among diverse organizations about the separation among deployed UAVs, we have defined this number of vehicles because it looks like representative of the environment scenario. The UAVs embed IEEE 802.11n WiFi technology and get a communication range of around 115~m. They fly 300~ft high to better reflect a more realistic permitted flight level for UAVs~\cite{faa2021}, with a constant speed of 20~m/s. All UAVs follow a 2D Random walk mobility model to collect environmental data. They collaboratively share their location to support building individual spatial awareness and update their positioning every 1.5~s. Although mobility models are not best suited for evaluating the proposed service performance in FANETs because they do not represent the natural behavior of the vehicles. Hence, we adopted the 2D Random walk model as a proof of concept. As each UAV has a limited operational time to cover the area adequately, these settings require a moderate to high density of vehicles in the target area. Therefore, FlySafe can support UAVs in monitoring the area continuously and reporting objects of interest found through the known close devices. For better analysis and comparisons, we have simulated this scenario with two distinct configurations, a BASELINE, with only honest UAVs, and a BASEATTK, with one MalUAV. Each simulation runs for 1200~s, and the applied metrics are averaged over 35 simulation runs to achieve a confidence interval of 95\%, except where otherwise described.

\begin{table}[!t]
	\centering
	\caption{Simulation settings}
        \relsize{-1.0}
	\begin{tabular}{m{4.0cm} m{4.0cm}<{\centering}}
		\toprule[1pt]
            \textbf{Setting} & \textbf{Value(s)} \\
		\midrule[0.5pt]
            Simulation time & 1200 s \\
		Map area & 1.5 km $\times$ 1.5 km \\  
            Flight level & 300 ft \\
		Number of UAVs & 40 \\
		Mobility Model & 2D Random Walk\\
            Flight speed & 20 m/s\\
		PHY/MAC Protocol & 802.11n\\
            Communication range & 115 m \\
		Propagation Model & Free-space\\
		Antenna Type & Omni-directional\\
            \# malicious node & 0, 1 \\
        \bottomrule[1pt]
	\end{tabular}
	\label{tab:simsettings}
\end{table}

\subsection{Metrics}

For supporting the evaluation, we have applied resilience and performance metrics to assess FlySafe's ability to provide and maintain a location service in the face of UAV mobility and against FDI attacks. For all metrics, we compute the average, the maximum, the~minimum, and the standard deviation achieved in all simulations. The assessment of resilience metrics enables us to point out the FlySafe behavior in providing the location service against disturbances caused by a MalUAV injecting false location data in the network. Hence, we measure the \textit{Age of Incorrect Information} (AoII), the \textit{Spatial Awareness Time} $\psi$, and the \textit{Age of Information} (AoI). We adapted the metric \textit{Age of Incorrect Information} (AoII)~\cite{maatouk2020age} to compute the period a UAV holds incorrect information about others. In this condition, even though a UAV receives location updates, its NL still remains incomplete. AoII equals the period of a UAV remains unaware about others, such that $\Gamma > 0$. The AoII value is defined as

\begin{equation}\label{eq:achieveAwareness}
AoII = \int_{0}^{\Tb} \Delta_{\lambda_{\scriptstyle{\{\Gamma \, > \, 0\}}}}(t) dt.
\end{equation}

We mainly focused on measuring the time a UAV keeps spatial awareness during operation. The \textit{Spatial Awareness Time} $\psi$ equals the total period a UAV successfully recognizes all others in its coverage area, i.e., the \textit{neighbor discovery error} is minimum ($\Gamma = 0$). The $\psi$ value is given by

\begin{equation}\label{eq:keepAwareness}
\psi = \mathlarger{\sum\limits_{t=0}^{\Tb}} \int_{0}^{\Tb} \Delta_{\lambda_{\scriptstyle{\{\Gamma \, = \, 0\}}}}(t) dt
\end{equation}

\noindent
where $\Delta_{\lambda_{\scriptstyle{\{\Gamma \, = \, 0\}}}}(t)$ is a time interval a UAV is successfully aware of its neighbors.

Lastly, we adapted the \textit{Tim e-Average Age} metric~\cite{yates2021age} to represent the \textit{Age of Information} (AoI) in the UAV network. AoI indicates UAV neighborhood information freshness and points out FlySafe behavior in providing and keeping UAV spatial awareness during all time slots $\Tb$. The instantaneous value of AoI is defined as

\begin{equation}\label{eq:aoi}
AoI = \int_{0}^{\Tb} \Delta(t) dt
\end{equation}

\noindent
where $\Delta(t)$ denotes the interarrival time and system time of a neighborhood information update.

The performance metrics aims to establish the ability of each UAV to achieve and keep an accurate perception of the spatial environment, i.e., spatial awareness. The assessment of data reliability in the location service follows four metrics: \textit{location error} $\Omega$, \textit{neighbor discovery error} $\Gamma$, \textit{convergence rounds} $\varphi$, and \textit{delay on location changes recognition} $\Upsilon$. As FlySafe runs in two distinct phases -- neighbor discovery and neighbor maintenance, we started by evaluating its accuracy to keep a correct perception of a neighbor UAV location, i.e., the \textit{location error} $\Omega$. The instantaneous $\Omega$ value at time slot $t$ is given by

\begin{equation}\label{eq:locError}
    \Omega^t = |d^r - d^m| 
\end{equation}

\noindent
where $d^r$ equals the distance of a neighbor to a UAV at the time it recognizes the neighbor's presence $t^r$, and $d^m$ means the distance to the new location due to UAV mobility during the time between neighbor UAV identifies itself through an \textit{Identification message} and $t^r$. 

We evaluate the neighbor discovery phase by determining the erroneous estimates in a time slot $t$, i.e., \textit{neighbor discovery error} $\Gamma$. Since UAV operation based on wrong knowledge about others is critical, $\Gamma$ minimization becomes essential to a safe flight. The instantaneous $\Gamma$ value at time slot $t$ is defined as

\begin{equation}\label{eq:neighError}
    \Gamma_{j}^t = W_{j}^t - W_{\Sb}^t 
\end{equation}

\noindent
where $W_{j}^t$ means the size of the neighborhood of a UAV $j$ in a time slot $t$. $W_{\Sb}^t$ is the size of the subset of UAVs in the simulation $\Sb \subset \Ub$ that should be within UAV $j$ coverage area in a time slot $t$, according to the network technology.

During the periods a UAV pursues spatial awareness, the \textit{convergence rounds} $\varphi$ mean the number of communication rounds, i.e., exchanged location per UAV required to achieve spatial awareness in each period. The $\varphi$ value is given by

\begin{equation}\label{eq:convRounds}
    \varphi = \sum_{\substack{t = 0 \\ \Gamma > 0}}^{t \leq T} (HM + IM + TM)
\end{equation}

\noindent
where HM, IM and TM correspond mean the number of \textit{Hello messages}, \textit{Identification messages}, and \textit{Trap messages}, respectively. As TM plays a pivotal role in the neighbor maintenance phase, we must minimize the \textit{delay on location changes recognition} $\Upsilon$, i.e., the period between the detection of a location change ($t_i^{d}$) by a UAV $i$ and the time a neighbor $j$ is aware of it ($t_j^{r}$). The instantaneous value of $\Upsilon$ is given by

\begin{equation}\label{eq:avgRec}
    \Upsilon = t_j^{r} - t_i^{d}. 
\end{equation}

\subsection{Results and Analysis}

In this section, we present and discuss the achieved results throughout the simulations in both evaluated scenarios -- BASELINE and BASEATTK. We analyze the location service provided by FlySafe mainly in the face of UAVs' mobility and its resilience against FDI attacks. 

\subsubsection{Resilience analysis}

Table~\ref{tab:aoii_aoi_spatial} shows UAVs keep spatial awareness for at least 81.5\% ($\psi =$ 978~s) of their operation time. However, $\psi$ eventually attains 1144.32~s, which means UAVs recognize their neighbors on 95.36\% of their flight time. Moreover, a MalUAV playing FDI attack (BASEATTK) slightly jeopardizes UAV spatial awareness. It causes a reduction of 7.7\% of UAV minimum $\psi$ value, 903~s. Still, the impact of the MalUAV on other operations is even smaller. In such conditions, UAVs keep spatial awareness commonly for 1129.85~s, i.e., a reduction of 1.2\% regarding BASELINE configuration. Such results points out that FlySafe provides a location service resilient to false locations shared on the network. Although UAVs achieve spatial awareness for a slightly shorter time than in BASELINE, they even fly aware of others during 94.15\% of their operation time.

\begin{table}[H]
    \centering
    \caption{Spatial awareness condition}
    \tabcolsep=0pt
    \relsize{-2}
    \begin{tabular}{m{1.9cm} m{0.1cm}<{\centering} m{0.8cm}<{\centering}m{1.2cm}<{\centering}m{0.8cm}<{\centering} m{0.1cm}<{\centering} m{0.6cm}<{\centering}m{1.2cm}<{\centering} m{0.1cm}<{\centering} m{0.6cm}<{\centering}m{1.0cm}<{\centering} m{0.7cm}<{\centering} m{0.1cm}<{\centering} m{0.8cm}<{\centering}m{0.8cm}<{\centering} m{0.1cm}<{\centering} m{0.7cm}<{\centering} m{0.9cm}<{\centering}m{1.2cm}<{\centering}}
 
	\hlineB{2}
        \textbf{Metric} & & \multicolumn{6}{c}{\textbf{AoI} (s)} & & \multicolumn{6}{c}{\textbf{AoII} (s)} & & \multicolumn{3}{c}{\textbf{$\psi$} (s)}\\ \cline{1-1} \cline{3-8}\cline{10-15}\cline{17-19}
            
        \textbf{Values}& & \multicolumn{3}{c}{\textbf{Averaged}} & & \multicolumn{2}{c}{\textbf{Sample}} & & \multicolumn{3}{c}{\textbf{Averaged}} & & \multicolumn{2}{c}{\textbf{Sample}} & & \multicolumn{3}{c}{\textbf{Averaged}}\\  \cline{1-1}\cline{3-5}\cline{7-8}\cline{10-12}\cline{14-15}\cline{17-19}
            
	\textbf{Config.} & & \textbf{min} & \textbf{max} & \textbf{avg} & & \textbf{min} & \textbf{max} & & \textbf{min} & \textbf{max} & \textbf{avg} & & \textbf{min} & \textbf{max} & & \textbf{min} & \textbf{max} & \textbf{avg}\\ \hline
  
        \textbf{BASELINE} && 7.51 & 310.50 & 53.68 && 0 & 1199.58 && 0 & 15.84 & \textcolor{blue}{1.57} && 0 & \textcolor{blue}{110.0} && 978 & 1198 & 1144.32 \\
           
        \textbf{BASEATTK} && 7.07 & 319.33 & 54.25 && 0 & 1199.80 && 0 & 18.94 & \textcolor{blue}{1.92} && 0 & \textcolor{blue}{131.8} && 903 & 1198 & 1129.85 \\
            
        \hlineB{2}
	\end{tabular}
	\label{tab:aoii_aoi_spatial}
\end{table}

Undoubtedly, UAVs alternate moments of unconsciousness and spatial awareness during operation due to their mobility, which causes frequent network disconnections. Further, a MalUAV by sharing a false location also compromises UAV spatial awareness. As shown in Table~\ref{tab:aoii_aoi_spatial}, UAVs commonly remain unaware of others for 1.57~s, but that condition can last for a maximum period of 9.1\% (AoII = 110~s) of their operation (BASELINE). In contrast, facing a MalUAV sharing false location (BASEATTK), AoII augments to 1.92~s, around 22\% compared to regular operation. Furthermore, UAVs remained unaware of their neighborhood for a maximum period of 11\% (AoII = 131.8~s) of their operation. Those values show that the service provided by FlySafe is resilient to a false location shared by MalUAVs, thus ensuring UAV's spatial awareness for up to 89\% of their operation time.

According to Table~\ref{tab:aoii_aoi_spatial} values, UAVs commonly keep spatial awareness (AoI) for at least 53.68~s when only their mobility triggers neighbor maintenance (BASELINE). Eventually, AoI extends for a maximum period of 310.50~s, whereas some UAVs attained an AoI of 1199.58~s. Such extreme conditions can take place in sparse environments where UAVs barely interact. However, UAVs typically get a MalUAV's presence in their coverage area. In this way, 
FlySafe's resilience to FDI attacks is expressed by an AoI similar to both scenarios. 

\subsubsection{Performance analysis}

We started evaluating the \textit{location error} $\Omega$ on neighbors' perception to build UAVs spatial awareness. As presented in Table~\ref{tab:location_delay}, a UAV typically identifies others with $\Omega$ of 0.87~m in a scenario free of threats (BASELINE). Under FDI attack, $\Omega$ almost doubles and attains 1.92~m. Therefore, although false location impacts FlySafe's performance, the location service faces MalUAV's operation with a low increase in $\Omega$ values. In such conditions, the sooner we identify a MalUAV, the less impact the false location has on the location service. Further, strategies like collision avoidance can consider $\Omega$ values to mitigate such threats and contribute to a safe flight.

\begin{table}[H]
    \centering
    \caption{Neighborhood perception}
    \setlength{\tabcolsep}{3.0pt}
    \relsize{-1.0}
    \begin{tabular}{lcccccccccccc}
    \hlineB{2}
    \textbf{Metric} & & \multicolumn{3}{c}{\textbf{$\Omega$} (m)} && \multicolumn{3}{c}{\textbf{$\Upsilon$} (ms)} & & \multicolumn{3}{c}{\textbf{$\Gamma$} (\# nodes)}\\ \cline{1-1}\cline{3-5}\cline{7-9}\cline{11-13}
		
    \textbf{Config.} & & \textbf{min} & \textbf{max} & \textbf{avg} & & \textbf{min} & \textbf{max} & \textbf{avg} & & \textbf{min} & \textbf{max} & \textbf{avg}\\ \hline
  
        BASELINE & & 0.16 & 1.14 & 0.87 & & 0.06 & 2845.7 & 2.7 &&0&2.25&0.38\\
        BASEATTK & & 0.93 & 2.67 & 1.92 & & 0.06 & 2920.7 & 2.7&&0&2.20&0.36\\

    \hlineB{2}
    \end{tabular}
    \label{tab:location_delay}
\end{table}

FlySafe ensures that UAVs recognize other location changes with similar delays ($\Upsilon$) for both configurations, as shown in the values in Table~\ref{tab:location_delay}. Typically, UAVs attain $\Upsilon$ of 2.7~ms. Since $\Upsilon$ relates to the time required for UAVs to recognize others' location changes, devices disregard the content of such information, whether it is true or not. Although the MalUAV injects false location data in BASEATTK configuration, FlySafe deals with the information regularly. Thus, any UAV performs similarly to share its location and recognize others' location changes. Further, FlySafe leverages all exchanged messages to update neighbors' locations. Therefore, even if location updates take place at every 1.5~s, the minimum $\Upsilon$ value attains 0.06~ms. Such value highlights the substantial improvement in UAVs' spatial awareness.

In a general way, as shown in Table~\ref{tab:location_delay}, FlySafe operates with low $\Gamma$ values in both evaluated scenarios, 0.38 and 0.36 to BASELINE and BASEATTK, respectively. Those values indicate that each UAV commonly discovers its neighborhood in any time slot with an error inferior to one UAV. Hence, UAVs typically successfully discover others, which highlights the advantage of opportunistic approaches to the location service.

\begin{figure}[!t]
\centering 
\includegraphics[width=2.4in]{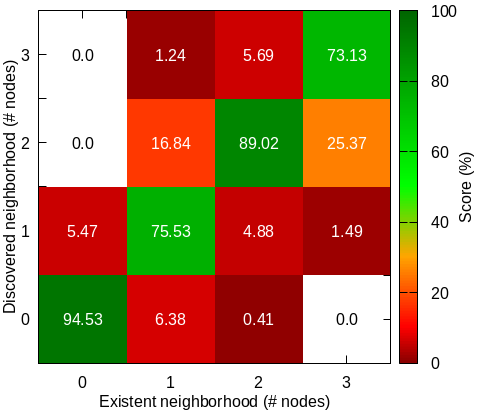}
\hspace{1cm}
\includegraphics[width=2.42in]{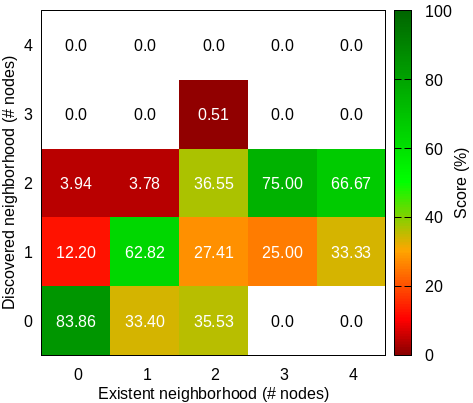}  \\

\hspace{-0.5cm} \footnotesize{(a) \hspace{7cm} (b)}

\caption{Spatial awareness condition during UAVs operation. (a) Honest UAV $\Gamma$ distribution. (b) Malicious UAV $\Gamma$ distribution.}
\label{fig:nd_error}
\end{figure}

We have selected one simulation run at random to analyze in detail $\Gamma$ values. Hence, we picked up the MalUAV and a couple of honest UAVs that interacted with it. Fig.~\ref{fig:nd_error}(a) showed $\Gamma$ distribution for an honest UAV with a maximum of three neighbors over the simulation. According to the hits on the main diagonal, the UAV commonly discovered its neighborhood. It achieved a maximum accuracy of 94.53\%. The false positives (FP) rates are over the main diagonal, so that they indicate that the device is aware of more UAVs than those in its coverage area. The UAV achieved a maximum FP rate of 16.84\%. In contrast, the values under the main diagonal are false negatives (FN) rates. In this way, they represent a severe threat to UAV operation since the UAV is aware of fewer UAVs than those in its coverage area. The FN rate achieved a maximum value of 25.37\%.

As shown in Fig.~\ref{fig:nd_error}(b), MalUAV behaves quite differently from the honest UAV. Although it partially discovers nearby devices, the hits mainly concentrate when there are few neighbors in its coverage area (83.36\%). Under these conditions, a UAV performs Neighbor discovery (ND) frequently, and the absence of nearby UAVs contributes to the success of building its spatial awareness. However, the increase in the number of neighbors jeopardizes ND. Since the MalUAV injects a false location on the network, the honest devices notify others about such behavior. Further, whether it behaves this way repeatedly, other UAVs segregate it from the network and disregard its messages. Thus, MalUAV interactions mainly take place with unknown devices, which leads to a high a high FN rate.

\begin{figure}[!t]
\centering 
\includegraphics[width=2.5in]{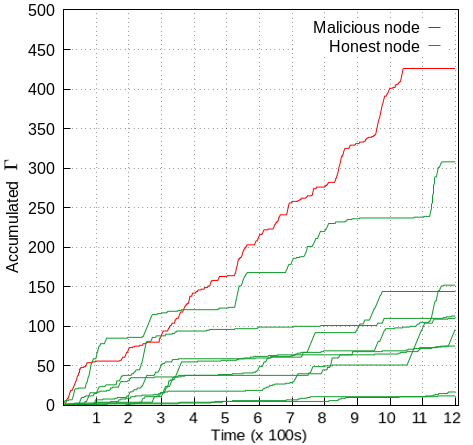}
\caption{UAVs spatial awareness behavior taking into account the accumulated $\Gamma$.}
\label{fig:nd_acc_error}
\end{figure}

The successfully maintenance of UAVs' spatial awareness happens when a UAV performs ND with a minimum error, i.e., $\Gamma = 0$. In this way, we accumulate the $\Gamma$ values of a couple of UAVs from the selected simulation run by summing such errors successively throughout their trajectory. Thus, we can represent their behavior in search of spatial awareness. As shown in Fig.~\ref{fig:nd_acc_error}, whenever $\Gamma = 0$, its accumulated value remains stable, which means the UAV is aware of others. It is worth noting that FlySafe ensures that honest UAVs maintain spatial awareness for long periods, alternating with brief moments without complete knowledge about their neighborhood. On the other hand, the increasing accumulated $\Gamma$ value for the MalUAV indicates that it behaves pretty unstable when interacting with its neighbors. Although FlySafe segregates it from the network, such a UAV eventually achieves spatial awareness during very short periods. Further, the stability of the accumulated $\Gamma$ value at the end of the MalUAV trajectory is due to the absence of other UAVs in its coverage area.

The analysis of \textit{convergence rounds} $\varphi$ evidences that FlySafe behaves accordingly in the simulated scenarios. As presented in Table~\ref{tab:rounds}, UAVs send 57.31 Hello messages (HM) to announce their presence and perform ND in the scenario free of threats (BASELINE). Generally, they attain $\varphi =$ 139.1 rounds in search of spatial awareness during their trajectory. In contrast, the MalUAV performing FDI attack increases $\varphi$ about 10\% higher than BASELINE. Such increment can be attributed mainly to the false location shared with other UAVs. It jeopardizes their spatial awareness, thus leading them to announce their presence more frequently. Hence, HM raises by 19\% from BASELINE to BASEATTK so that honest UAVs improve neighbor discovery and maintenance by exchanging more messages ($\varphi$ = 154.21) to achieve spatial awareness. 

\begin{table}[!htb]
	\centering
	\caption{Convergence rounds to achieve spatial awareness}
        \setlength{\tabcolsep}{9pt}
	\renewcommand{\arraystretch}{1.4}
        \relsize{-1.0}
	\begin{tabular}{llcccc}
        \hlineB{2}
	\multirow{2}{*}{\textbf{Config.}} & \multirow{2}{*}{\textbf{Event}} &\multicolumn{3}{|c}{\textbf{\# rounds}}& \multirow{2}{*}{\textbf{$\varphi$}} \\ \cline{3-5} 
        && HM & IM & TM & \\ \hline
        \multirow{2}{*}{\textbf{BASELINE}} & Send& \textcolor{blue}{57.31} & 6.99 & 24.1 & \multirow{2}{*}{\textcolor{blue}{139.1}}\\
        &Receive& 7.17 & 17.78 & 25.75 & \\ \hline
            
        \multirow{2}{*}{\textbf{BASEATTK}} & Send& \textcolor{blue}{68.23} & 7.84 & 26.86 & \multirow{2}{*}{\textcolor{blue}{154,21}}\\
            & Receive& 8.05 & 16.76 & 26.47 & \\
        
        \hlineB{2}
	\end{tabular}
	\label{tab:rounds}
\end{table}

\section{Conclusion} \label{sec:conclusion}
In this work, we proposed FlySafe, a resilient UAVs location sharing service based on information freshness and opportunistic deliveries to enable UAVs to perform a safe flight in order to provide an application service. Joining opportunistic approaches like direct delivery and crowdsourcing allows UAVs to timely deliver their location actively. The age of UAVs' location information triggering devices discovery leverages the freshness of UAVs location to improve overall spatial awareness. Since FlySafe segregates from the network UAVs performing false data attacks, the influence of false locations on UAVs' spatial awareness becomes negligible. Simulation results validate the effectiveness of the proposed service. FlySafe keeps UAVs' spatial awareness for most of their operation time. Despite a false location, UAVs recognize location updates and discover others like in the face of true information, evidencing that FlySafe is resilient to false locations injected into the network. 

\section*{Acknowledgment}

This work was supported by the National Council for Scientific and Technological Development (CNPq/Brazil), grant number 307752/2023-2.

\bibliographystyle{elsarticle-num} 
\bibliography{references}

\end{document}

%% file: mypackages.tex
\usepackage{scrextend}
\usepackage{mathtools}
\usepackage{wrapfig}
\usepackage{graphicx,url}
\usepackage{booktabs}
\usepackage[utf8]{inputenc}
\usepackage{hhline}
\usepackage{enumitem, array}
\usepackage{boldline}
\usepackage{xurl} 
\usepackage{relsize}
\usepackage{setspace}
\usepackage[T1]{fontenc} 
\usepackage{arydshln}
\usepackage{hyperref}

\usepackage{mathtools}
\usepackage{graphicx}
\usepackage{multirow} 

\usepackage{moresize} 

\usepackage[english,ruled,linesnumbered]{algorithm2e} 

\SetCommentSty{mycommfont}

\usepackage{indentfirst}
\usepackage{verbatim}
\usepackage{amsmath}
\usepackage{amssymb} 
\usepackage{amsfonts}
\usepackage{array}
\usepackage{listings}

\usepackage{xcolor}

\usepackage[export]{adjustbox}
\usepackage{float}
\usepackage[normalem]{ulem} 
\usepackage{ascii,pifont,bbding,wasysym,ifsym,dictsym,keyval}
\usepackage{pifont}
\usepackage{lscape} 

\usepackage{tabularx,colortbl} 

\usepackage{calrsfs} 
\DeclareMathAlphabet{\pazocal}{OMS}{zplm}{m}{n}
\newcommand{\Lb}{\pazocal{L}}

\newcommand{\Ub}{\pazocal{U}}
\newcommand{\Tb}{\pazocal{T}}
\newcommand{\Wb}{\pazocal{W}}
\newcommand{\Pb}{\pazocal{P}}
\newcommand{\Sb}{\pazocal{S}}

\newcommand{\Ua}{\mathcal{U}}
\newcommand{\Wa}{\mathcal{W}}
\newcolumntype{L}[1]{>{\raggedright\let\newline\\\arraybackslash\hspace{0pt}}m{#1}}
\newcolumntype{C}[1]{>{\centering\let\newline\\\arraybackslash\hspace{0pt}}m{#1}}
\newcolumntype{R}[1]{>{\raggedleft\let\newline\\\arraybackslash\hspace{0pt}}m{#1}}

\usepackage{tikz}


%% file: mycommands.tex
\let\oldnl\nl
\newcommand{\nonl}{\renewcommand{\nl}{\let\nl\oldnl}}

\definecolor{color4}{RGB}{179, 43, 59}
\definecolor{LightGray}{RGB}{236, 236, 236}

